%% file: paper.tex
\documentclass[twocolumn,showpacs,aps,prl,preprintnumbers,superscriptaddress]{revtex4}

\usepackage{xspace,graphicx,dcolumn,amsmath,epsfig}

\input babarsym.tex

\input mysymbols.tex

\newcommand{\BABARPubYear}     {06}
\newcommand{\BABARPubNumber}  {009}

\newcommand{\SLACPubNumber} {11972}

\newcommand{\gevccc}{\ensuremath{{\mathrm{\,Ge\kern -0.1em V^2\!/}c^4}}\xspace}

\def\figurebox#1#2#3{%
    \def\arg{#3}%
    \ifx\arg\empty
    {\hfill\vbox{\hsize#2\hrule\hbox to #2{\vrule\hfill\vbox to #1{\hsize#2\vfill}\vrule}\hrule}\hfill}%
    \else
    {\hfill\epsfbox{#3}\hfill}%
    \fi}

\long\def\inst#1{\par\nobreak\kern 4pt\nobreak
    {\it #1}\par\vskip 10pt plus 3pt minus 3pt}

\begin{document}


\preprint{\babar-PUB-\BABARPubYear/\BABARPubNumber, SLAC-PUB-\SLACPubNumber} 

\title{
{\Large \bf \boldmath
Search for the reactions \eemtau\ and \eeetau} 
}

\input authors_jul2006.tex

\date{\today} 

\begin{abstract}

We report on a search for the lepton-flavor-violating processes
$e^+e^- \rightarrow \mu^+\tau^-$ and $e^+e^- \rightarrow
e^+\tau^-$. The data sample corresponds to an integrated luminosity of
211 fb$^{-1}$ recorded by the \babar\ experiment at the SLAC PEP-II
asymmetric-energy \B\ Factory at a center-of-mass energy of $\sqrt{s}
= 10.58$ GeV. We find no evidence for a signal and set the 90\%
confidence level upper limits on the cross sections to be
$\sigma_{\mu\tau} < 3.8$ fb and $\sigma_{e\tau} < 9.2$ fb.  The ratio
of the cross sections with respect to the dimuon cross section are
measured to be $\sigma_{\mu\tau}/\sigma_{\mu\mu}<3.4\times10^{-6}$ and
$\sigma_{e\tau}/\sigma_{\mu\mu}<8.2\times10^{-6}$.

\end{abstract}


\maketitle


Within the Standard Model (SM), the fermion mass matrices and the
mechanism of electroweak symmetry breaking remain unexplained.
Lepton-flavor is not a conserved quantity protected by an established
gauge principle.  Extensions to the SM which include our
knowledge of neutrino masses and mixing~\cite{superk} predict
lepton-flavor-violation (LFV) at a level many orders of magnitude
below the current experimental sensitivity~\cite{susy}.

Searches for LFV have primarily concentrated on the decay of the
lepton.  Limits in a number of muon decay channels have reached the
$10^{-11}-10^{-12}$ level~\cite{pdg} while recent measurements of LFV
in tau decays have placed limits on the branching fractions ${\cal
B}(\tau^{\pm}\rightarrow \mu^{\pm}\gamma) < 6.8\times 10^{-8}$ and
${\cal B}(\tau^{\pm}\rightarrow e^{\pm}\gamma) < 1.1\times
10^{-7}$~\cite{mugamma} at the 90\% confidence level (CL).

There are theories that suggest lepton-flavor can be
conserved in lepton decay but still be present in production. Some of
these models allow for channels such as $e^+e^- \rightarrow
\mu^+\tau^-$ and $e^+e^- \rightarrow e^+\tau^-$ through the $Q^2$
evolution of the off-diagonal elements of the fermion mass
matrices~\cite{bordes}.  Experimental limits on LFV in
production are considerably weaker than for decay. At center-of-mass (CM)
energies, $\sqrt{s} =29$\gev, there are limits on the cross section
ratios $\sigma_{\mu\tau}/\sigma_{\mu\mu}< 6.1\times 10^{-3}$ and
$\sigma_{e\tau}/\sigma_{\mu\mu}< 1.8\times 10^{-3}$ (95\%
CL)~\cite{markII}; at $\sqrt{s} = 92$\gev, where $Z^0$ exchange
dominates, ${\cal B}(Z^0\rightarrow \mu\tau, e\tau) < {\cal O} (1)
\times 10^{-5}$ (95\% CL)~\cite{lep}. The best limits from searches at
LEP energies above the $Z^0$ peak are $\sigma_{\mu\tau}<64$\fb and
$\sigma_{e\tau}<78$\fb (95\% CL)~\cite{opal2}. No equivalent
measurements exist at the lower energies accessible by the \babar\
detector.

We present results on two modes of the process \eeltau, where $l^+$ is
an electron or muon and the $\tau^-$ decays either to $\pi^- \pi^+
\pi^- \nu_\tau$ or $\pi^- \nu_\tau$, using data recorded by the
\babar\ detector at the SLAC \pep2\ asymmetric-energy \epem\ storage
rings. Inclusion of the charge-conjugate reaction $e^+e^- \rightarrow
l^-\tau^+$ is assumed throughout this paper.  The data sample
corresponds to an integrated luminosity of \L = \lumion\ recorded at a
CM energy of $\sqrt{s}=10.58\gev$.

The \babar\ detector is described in detail in Ref.~\cite{detector}.
Charged particles are reconstructed as tracks with a 5-layer silicon
vertex tracker and a 40-layer drift chamber (DCH) inside a 1.5 T
solenoidal magnet.  An electromagnetic calorimeter (EMC) is used to
identify electrons and photons.  A ring-imaging Cherenkov detector
(DIRC) is used to identify charged hadrons and provides additional
electron identification information. Muons are identified by an
instrumented magnetic-flux return (IFR).

Monte Carlo (MC) simulation is used to evaluate the background
contamination and selection efficiency. The simulated backgrounds are
also used to cross-check the selection optimization procedure and for
studies of systematic effects; however, the final background yield
estimation relies solely on data. The signal $e^+e^- \rightarrow
l^-\tau^+$ channels are simulated using EvtGen~\cite{evtgen} in which
photon radiation is handled by the PHOTOS package~\cite{photos} to an
accuracy better than 1\%. The background $\tau$-pair events are
simulated using the \kk\ MC generator~\cite{kk}.  The $\tau$ decays
are modeled with Tauola~\cite{tauola} according to measured rates with
the decay \hhhnu\ assuming an intermediate $a^-_1(1260)$ axial-vector
state~\cite{pdg,cleoopal}. We also generate light quark continuum
events ($e^+e^- \rightarrow \qqbar,$\ $q=u,d,s$), charm, dimuon,
Bhabhas, \BB\ and two-photon events~\cite{evtgen,mcgen}. The detector
response is simulated with GEANT4~\cite{geant4} and all simulated
events are reconstructed in the same manner as data.

The signature of the signal process in the CM frame is 
an isolated high-momentum muon or electron recoiling against either
one or three charged pions and no neutral particles. The reconstructed
mass of the missing neutrino should be consistent with a massless
particle and the invariant mass of the recoiling pions and neutrino
consistent with that of the $\tau$.

We search for events with zero total charge and either two or four
well-measured charged tracks originating from the $e^+e^-$ interaction
region. All charged tracks must be isolated from neutral energy
deposits in the EMC and be within the acceptance of the EMC, DIRC and
IFR to ensure good particle identification. One track must be
identified as either an electron or muon with a CM momentum greater
than 4.68\gevc\ and no other track identified as a kaon or lepton. The
electron momentum is corrected for energy loss from Bremsstrahlung
emission by including in the electron momentum the energies of
isolated calorimeter deposits consistent with a photon within a cone
of radius 0.1 \rad\ around the initial track momentum vector.

In the CM system, the event topology must be consistent with an
$e^+/\mu^+$ recoiling against the remaining tracks. We calculate the
thrust axis~\cite{thrust} using all the charged and neutral deposits in
the event and define two hemispheres with respect to the plane normal
to the thrust axis and require that the $e^+/\mu^+$ and the
other tracks to be in separate hemispheres. 

The $\tau$ has a fixed CM energy and momentum:

\begin{equation}
\label{eq:ktau}
E^*_\tau = \frac{\sqrt{s}}{2} + \frac{(M^2_{\tau}  - M^2_l)}{2\sqrt{s}}, \hspace{0.25cm}
\mid {\bf p^*_\tau}\mid = \sqrt{ E_\tau^{*2} - M^2_{\tau}}
\end{equation}


\noindent where $M_{\tau}$ and $M_l$ are the masses of the $\tau$ and
$e^+/\mu^+$, respectively~\cite{pdg}.  We define the direction of the
$\tau$ as opposite to that of the $e^+/\mu^+$ and assign it the
momentum from Equation~\ref{eq:ktau}. The CM four-momentum of the
missing neutrino from the $\tau$ decay, $p^*_\nu$, is defined as
$p^*_\tau - \pres$, where \pres\ is the sum of the CM four-momenta of
the pions. The reconstructed $\tau$ mass is defined to be $m_\tau =
\sqrt{(\eres + \mid {\bf p^*_\nu}\mid )^2 - \mid {\bf p^*_\tau}\mid^2
}$ where \eres\ is the CM energy of the pions.

Events are rejected if the quantity \DeltaE, the difference between
the $e^+/\mu^+$ CM energy and $\sqrt{s}/2$, is less than $-0.5$\gev or
greater than 0.2\gev.  True signal events will have \DeltaE\ $\sim
-0.15$\gev while $e^+e^-\rightarrow\mu^+\mu^-$ or 
$e^+e^-\rightarrow e^+e^-$ events will peak
at zero and $e^+e^-\rightarrow\tau^+\tau^-$ background events have large negative
\DeltaE.  The \DeltaE\ resolution is approximately 50\mev. Events with
converted photons are also rejected, where a converted photon is
defined to be a pair of oppositely charged tracks assumed to have the
electron mass and coming from a vertex with a combined mass less than
$150\mevcc$.

We use a number of kinematic variables to suppress backgrounds.  The
missing event energy in the CM frame, \emiss, defined as the
difference between $\sqrt{s}$ and the sum of the charged track
energies, is distributed uniformly for signal but peaks at zero or
near $\sqrt{s}/2$ for the most important backgrounds. The missing mass
squared, \mmiss, should be consistent with zero. A requirement on the
maximum neutral energy cluster in the detector, $E_\gamma$, eliminates
events with neutral pions or photons~\cite{taunu}. 
A requirement on the angle in
the CM between the direction of the neutrino and the beam axis in the
$e^-$ beam direction, $\cos^*(\theta_\nu)$, ensures the reconstructed
neutrino is within the detector acceptance to reject events
with significant radiation along the beam direction. The angle in the
CM between the direction of the neutrino and the $\tau$,
$\theta^*_{\tau\nu}$, is used to reject background events with a
back-to-back track topology such as dimuon and Bhabha production. An
event is accepted if it falls within a two-dimensional region defined
with respect to $m_\tau$ and the $e^+/\mu^+$ CM momentum, $p^*_l$.
Events in this region are then used in a maximum likelihood fit to
extract the signal yield.

\begin{table}[ht!]
\caption{Selection criteria for the decay modes. 
The same criteria are used for the $e^+$ and 
$\mu^+$ lepton flavors except for \emiss.}
\begin{ruledtabular}
\begin{tabular}{lrclrcl}
            & \multicolumn{3}{c}{ \eeltau }
            & \multicolumn{3}{c}{ \eemtau\ ($e^+\tau^-$) } \\ 
            & \multicolumn{3}{c}{ \hhhnu  }
            & \multicolumn{3}{c}{ \hnu }      \\ \hline
\emiss (\gev)            &  0.015 &$-$& 3.23  & 0.65   &$-$& 4.55 (4.0) \\
\mmiss (\gevccc)         &        &$<$& 0.56  &        &$<$& 0.65       \\
$E_{\gamma}$ (\gev)      &        &$<$& 0.20   &       &$<$& 0.15       \\
$\cos^*(\theta_\nu)$       & $-$0.9 &$-$& 0.9   & $-$0.9 &$-$& 0.7        \\
$\theta^*_{\tau\nu}$ &        &$>$& 0.015 &        &$>$& 0.090       \\
$m_\tau$  (\gevcc)       &    1.6 &$-$& 2.0   &    1.6 &$-$& 2.0        \\
$p^*_{l}$  (\gevc)       &    4.90 &$-$& 5.32 &   5.02 &$-$& 5.32       \\
\end{tabular}
\end{ruledtabular}
\label{tab:sel2}
\end{table}

The values of the selection criteria are shown in
Table~\ref{tab:sel2}. We optimize the selection sensitivity by
defining a nominal signal box with a width of three standard
deviations in the reconstructed $m_\tau$ and $p^*_l$. The resolutions
on $m_\tau$ and $p^*_l$ are approximately $10\mevcc$ and $45\mevc$,
respectively. The values of the selection criteria are chosen to
maximize the discriminant $S/\sqrt{B}$ where $S$ is the number of MC
signal events in the nominal signal box and $B$ is the number of data
events accepted outside this region but within $1.5< m_{\tau} <
2.2$\gevcc\ and the $p^*_l$ boundaries given in
Table~\ref{tab:sel2}. We repeated the procedure using background MC
within the nominal signal box instead of data and this produced
consistent results. The signal MC reconstruction efficiencies and
their statistical error after the application of these selection
criteria are shown in Table~\ref{tab:results1}.


The backgrounds are dominated by $e^+e^-\rightarrow\tau^+\tau^-$
decays where one $\tau$ decays to an $e^+/\mu^+$ plus neutrinos and
the other to either $\pi^- \pi^+ \pi^- \nu_\tau$ or $\pi^-
\nu_\tau$. Light quark continuum processes are predicted to contribute
significantly to \amu\ only and events from
$e^+e^-\rightarrow\mu^+\mu^-$ are only present in \pimu.  Charm and
\BB\ backgrounds are eliminated by the track multiplicity and
\DeltaE\ requirement and all other backgrounds are negligible.

An extended unbinned maximum likelihood (ML) fit to the variables
$m_{\tau}$ and $p^*_l$ is used to extract the total number of signal
and background events separately for each mode. The likelihood
function $L$ is:

\begin{equation}
L = \frac{e^{-\sum_j n_j}}{N!}\prod_i^N \sum_j n_j{\cal P}_j(\vec{x}_i)
\end{equation}

\noindent where $n_j$ is the yield of events of hypothesis $j$ (signal
or background) and $N$ is the number of events in the sample. 
The individual background components comprise
$e^+e^- \rightarrow \tau^+\tau^-$, $e^+e^- \rightarrow
\mu^+\mu^-(\gamma)$ and light quark continuum decay modes. ${\cal
P}_j(\vec{x}_i)$ is the corresponding probability density function
(PDF), evaluated with the variables $\vec{x}_i=\{m_{\tau},p^*_l\}$ for
the $i$th event. For the signal, we use double Crystal Ball
functions~\cite{crystal} for both $m_{\tau}$ and $p^*_l$. Due to
correlations between $m_{\tau}$ and $p^*_l$ for non-signal events, we
use a two-dimensional non-parametric PDF obtained from
MC for the backgrounds~\cite{parametric}. In the maximum likelihood
fit to the data, the parameters of the PDFs are fixed to the values
determined from MC and only the signal and the background component
yields are allowed to float. The statistical errors on the yields 
by the ML fit are roughly a factor of two smaller than 
those achievable with a simple counting experiment.

We check the robustness of the fitting procedure against variations in
  the signal size and background shape.  We first fit the data outside
  the signal region with the MC background PDFs only, to determine
  their amplitudes.  Using these PDFs for the background, we generate
  trial distributions including a Poisson-distributed number of
  simulated signal events, and perform the fit for each.  We use 1000
  trials at each of twenty values of the average signal yield between
  0 and 100 events, and find the fitted signal yield to be unbiased
  and the statistical uncertainty to be estimated correctly.
  Secondly, we generate a set of trial distributions in which the
  relative amplitudes of the simulated background components are
  changed, and confirm that this does not bias the fitted signal
  yield. 

As a validation check, we compare the predicted MC background levels
and distributions of the variables from Table~\ref{tab:sel2} to the
data in the region outside the nominal signal box and find that they
are in agreement. We also extrapolate the fitted background PDFs from
the region outside the nominal signal region into the nominal signal
region and predict (measure) $193\pm9$ (202) and $143\pm7$ (154) for
\amu\ and \pimu, respectively, and $112\pm7$ (128) and $90\pm6$ (75)
events for \ael\ and \piel, respectively, where the error is
statistical only.  The predicted and measured values are consistent
within the statistical errors.

From the reconstructed MC efficiency, we can estimate the predicted
number of background events and compare to the results of the ML
fit. For $e^+e^-\rightarrow\tau^+\tau^-$, the predicted (ML fitted)
background in the fitted region is $750\pm43$ ($775\pm19$) and
$494\pm40$ ($385\pm35$) events for \amu\ and \pimu, respectively, and
$414\pm41$ ($518\pm41$) and $319\pm45$ ($331\pm18$) events for \ael\
and \piel, respectively.  The dimuon background to \pimu\ is predicted
(ML fitted) to be $114\pm38$ ($189\pm30$). For the light continuum
background, the MC predicts (ML fitted) $119\pm24$ ($129\pm40$) and
$19\pm9$ ($18\pm35$) events for \amu\ and \ael, respectively.  The
predicted and fitted values agree within errors.

The main sources of systematic error on the signal yield come from
uncertainties in the reconstruction, the \hhhnu\ decay mechanism and
the fit procedure. A relative systematic uncertainty of 0.8\% per
track, added linearly for all charged tracks in the event, is applied
to account for differences in MC and data charged particle
reconstruction. A relative systematic uncertainty of 1.0\% per charged
pion track and 1.3\% per $e^+/\mu^+$ track, added linearly for each
charged track, is applied to account for differences in MC and data
particle identification efficiencies.

A possible non-axial-vector decay mechanism for the decay \hhhnu\  
is not completely ruled out by current measurements~\cite{pdg}.  To
estimate this effect, the signal MC events were generated with 90\%
axial-vector and 10\% phase-space decays and the difference in the
reconstruction efficiency compared to 100\% $a^-_1(1260)$ decays
applied as a systematic. This introduces a relative systematic
uncertainty of 3.2\%.

The largest systematic error come from the variation of the PDF fit
parameters within their fitted errors. The two-dimensional
non-parametric background PDFs show small structures that depend on MC
statistics and the value of the smoothing parameter
used~\cite{parametric}. By varying the smoothing parameter,
using different functional forms and varying the fitted parameters
within their uncertainties, we derive a systematic error of $\sim0.5$
events. To investigate possible mismodeling of the detector acceptance 
and response, we repeat the analysis with each selection criterion
varied by the resolution on the corresponding variable. All changes to
the signal yield are smaller than the statistical error and we
conservatively take the largest change in each case as a systematic
uncertainty, which ranges from 2.5 to 4.4 events.
The total systematic error is between 2.6 and 4.4 events
and our final limit on the cross sections is dominated by the
statistical error which is of the order of 10 events.

\begin{figure}[ht!] \begin{center} 
\begin{tabular}{c}
    \epsfig{file=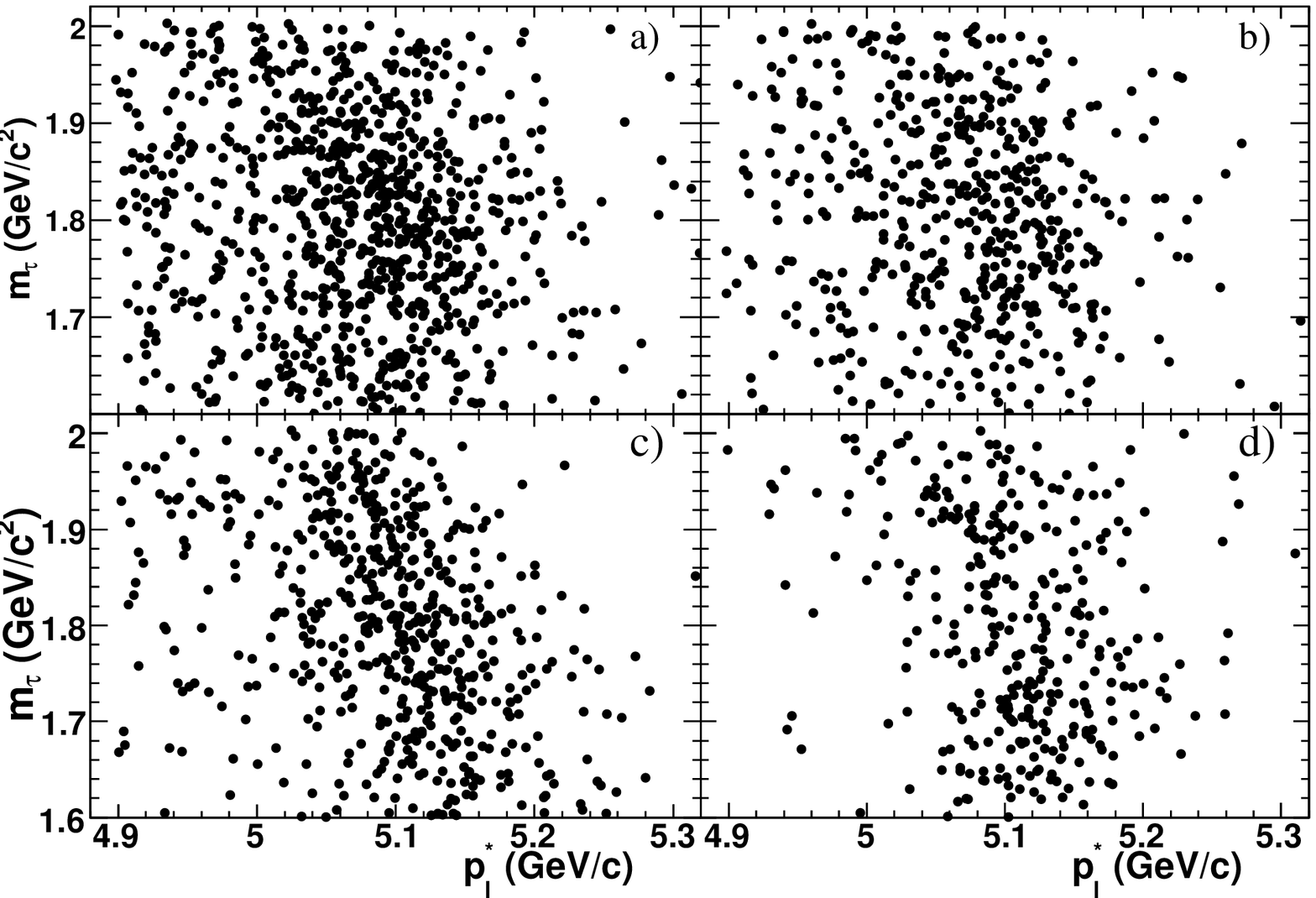,width=1.0\columnwidth}
\end{tabular}
\caption{$m_\tau$ versus \plep\ for reconstructed candidates for: 
a) \amu; b) \pimu; c) \ael; and d) \piel.}
\label{fig:scatter}
\end{center} \end{figure}
\begin{figure}[ht!] \begin{center} 
\begin{tabular}{c}
    \epsfig{file=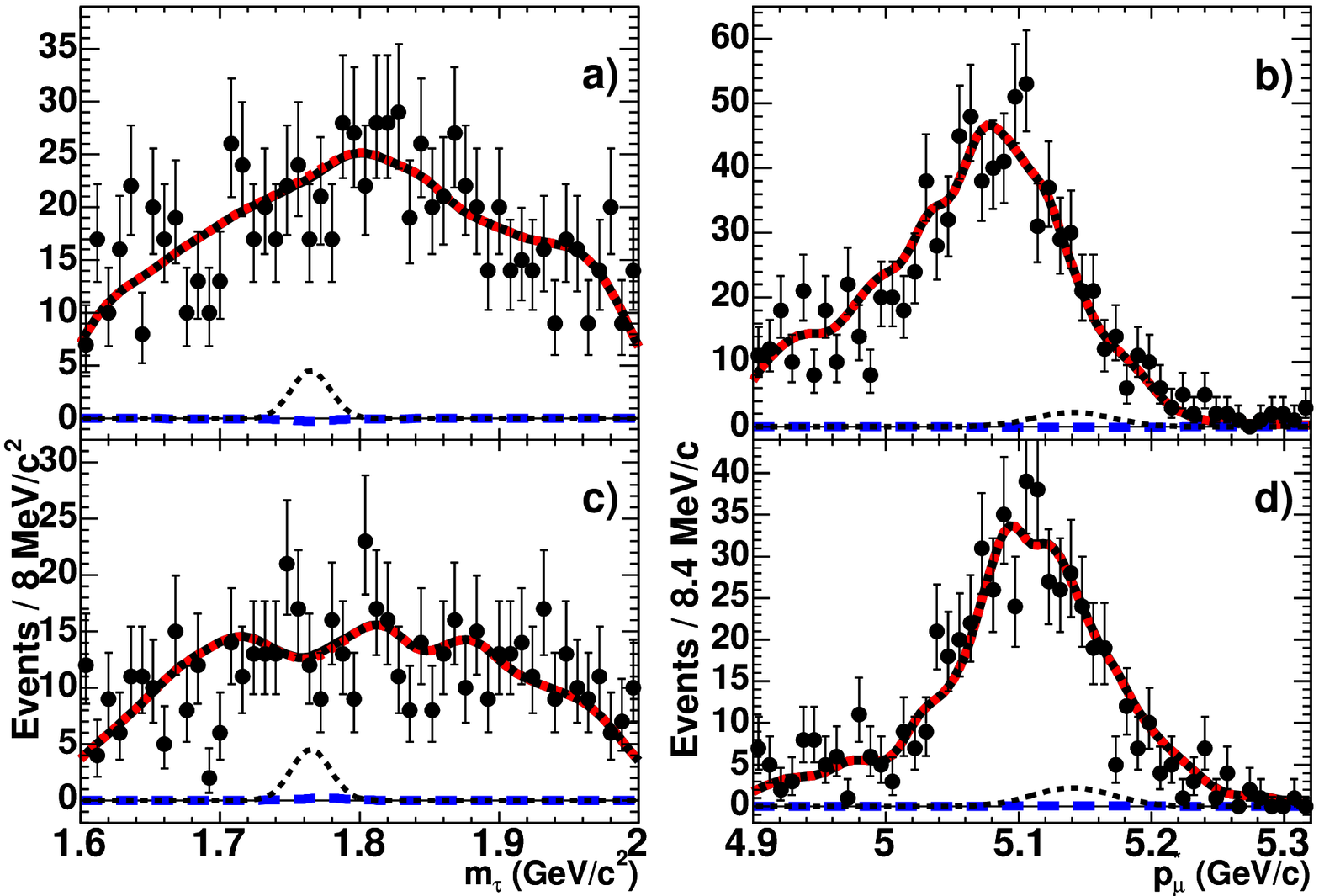,width=1.0\columnwidth}
\end{tabular}
\caption{Reconstructed distributions for 
\eemtau\ candidates: a) $m_\tau$ and b) \plepmu\ for \hhhnu; 
and c) $m_\tau$ and d) \plepmu\ for \hnu. 
The projection of the ML fit (solid line) hides the 
background component (dotted line). The projection of the few signal
events is shown on the horizontal axis as a dashed line. The peaking
dotted line shows the expected MC signal distribution at the 90\% CL
upper limit.}
\label{fig:a1}
\end{center} \end{figure}
\begin{figure}[ht!] \begin{center} 
\begin{tabular}{c}
    \epsfig{file=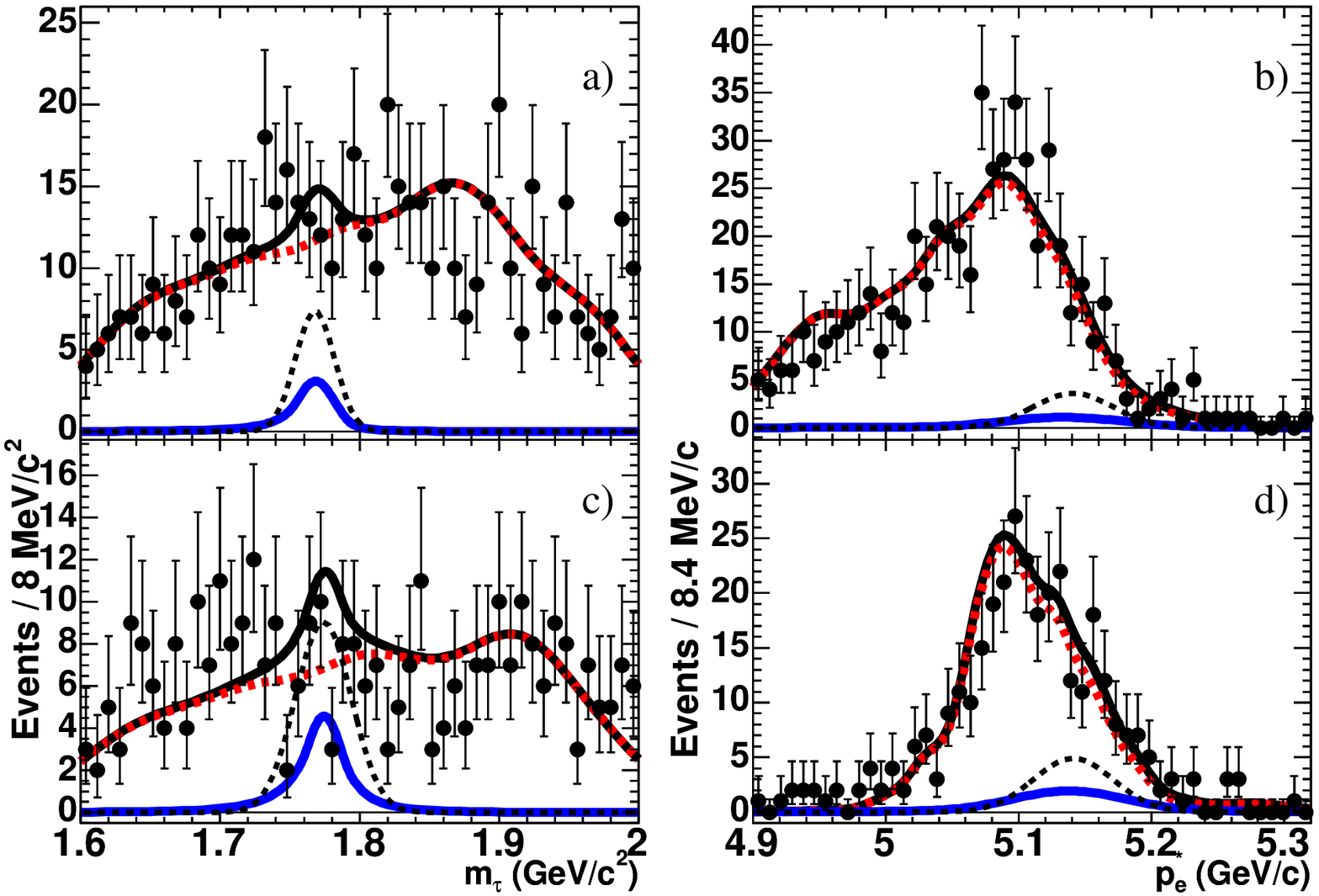,width=1.0\columnwidth}
\end{tabular}
\caption{Reconstructed distributions for 
\eeetau\ candidates: a) $m_\tau$ and b) \plepe\ for \hhhnu; 
and c) $m_\tau$ and d) \plepe\ for \hnu. 
The solid line is the projection of the ML fit, 
the dotted line is the background component and the dashed line is the 
signal component. The peaking
dotted line shows the expected MC signal distribution at the 90\% CL
upper limit.}
\label{fig:pi}
\end{center} \end{figure}

The $m_\tau$ and \plep\ distributions for the modes are shown in
Figure~\ref{fig:scatter} and the projections are
shown in Figures~\ref{fig:a1} and~\ref{fig:pi}. The projection of the
signal PDF is shown as the dashed line, the background PDFs as the
dotted line and the total PDF as the solid line. 
The central value of the cross section for \eeltau\ is given by $\sigma =
N/\eta\epsilon{\cal L}$ where $N$ is the number of signal events,
$\eta$ the signal reconstruction efficiency and $\epsilon$ is the
\hhhnu\ or \hnu\ branching fraction. The measurements are not
statistically different from the null hypothesis and we obtain 90\% CL
upper limits by finding the maximum number of signal events $N$ such
that the integral of the total likelihood function is 90\% of the
total integral. From MC studies~\cite{kk}, the total cross section of
the process $e^+e^-\rightarrow \mu^+\mu^-$ at $\sqrt{s} = 10.58\gev$
is $\sigma_{\mu\mu} = (1.13\pm0.02)$\nb and we use this to calculate
90\% CL upper limits on the ratio of the cross sections with respect
to the dimuon cross section.  The central values of the signal yields
from the maximum likelihood fit and the upper limits on the cross
sections and cross section ratios are given in
Table~\ref{tab:results1}.

\begin{table}[ht!]
\caption{Summary of the signal yields, 
cross sections and ratios of cross sections to 
dimuon cross section. The first uncertainty is the
statistical error and the second systematic.}
\begin{ruledtabular}
\begin{tabular}{lll}
\eemtau\             & \hhhnu & \hnu \\ \hline
Total Events  & 905 & 575 \\
Signal Events & $-1.37\pm9.9\pm2.6$    & $1.9\pm 10.1 \pm 4.4$\\
Signal Events (\ulimit\ CL) & $<19.2$    & $<19.9$\\
MC Efficiency (\%) & $18.5\pm0.2$ & $9.62\pm0.14$ \\
$\sigma_{\mu\tau}$ (\fb)& $-0.35\pm2.6\pm0.7$ & $0.85\pm 4.5 \pm\ 2.0$\\
$\sigma_{\mu\tau}$ (\ulimit\ CL) & $<\amuxstr$\fb & $<\pimuxstr$\fb\\ 
$\sigma_{\mu\tau}$ (95\% CL) & $<5.91$\fb & $<11.4$\fb\\ 
$\sigma_{\mu\tau}/\sigma_{\mu\mu}$ (\ulimit\ CL)      & $<\amubftr$ & $<\pimubftr$ \\ 
$\sigma_{\mu\tau}/\sigma_{\mu\mu}$ (95\% CL) & $<5.2\times10^{-6}$ & $<10.1\times10^{-6}$ \\ \hline
& & \\ \hline
\eeetau       & \hhhnu & \hnu \\ \hline
Total Events  & 537 & 332 \\
Signal Events & $15.9\pm10.3\pm2.7$ &  $10.7\pm8.8\pm2.7$\\
Signal Events (\ulimit\ CL) & $<32.3$    & $<25.8$\\
MC Efficiency (\%) & $11.73\pm0.15$ & $11.9\pm0.15$ \\
$\sigma_{e\tau}$ (\fb)& $6.5\pm4.2\pm1.1$ &  $3.9\pm 3.2\pm1.0$\\
$\sigma_{e\tau}$ (\ulimit\ CL) & $<\aelxstr$\fb & $<\pielxstr$\fb\\ 
$\sigma_{e\tau}$ (95\% CL) & $<14.8$\fb & $<11.1$\fb\\ 
$\sigma_{e\tau}/\sigma_{\mu\mu}$ (\ulimit\ CL) & $<\aelbftr$ & $<\pielbftr$\\
$\sigma_{e\tau}/\sigma_{\mu\mu}$ (95\% CL) & $<13.1\times10^{-6}$ & $<9.8\times10^{-6}$\\
\end{tabular}
\end{ruledtabular}
\label{tab:results1}
\end{table}

We combine the \hhhnu\ and \hnu\ decays and calculate 90\% CL upper
limits on the cross sections of $\sigma_{\mu\tau}<\muxstr$\fb for
\eemtau\ and $\sigma_{e\tau}<\elxstr$\fb for \eeetau. The 90\% CL
upper limits on the ratio of the cross sections with respect to the
dimuon cross section are calculated to be
$\sigma_{\mu\tau}/\sigma_{\mu\mu}<\mubftr$ and
$\sigma_{e\tau}/\sigma_{\mu\mu}<\elbftr$.  For comparison with
previous LEP results measured at $\sqrt{s} \geq 92$\gev, 
the 95\% CL upper limits on the cross sections and
ratio of cross sections are $4.6$\fb and $4.0\times10^{-6}$ for
\eemtau\ and $10.1$\fb and $8.9\times10^{-6}$ for \eeetau,
respectively.

In conclusion, we have performed the first search at a CM energy of
$\sqrt{s} = 10.58\gev$ of the lepton-flavor-violating production
processes \eemtau\ and \eeetau.  No statistically significant signal
events were observed in any of the decay modes. Upper limits have been
placed on the cross sections and ratios of cross sections to the
dimuon cross section to form limits on \eemtau\ and \eeetau.

\input acknow_PRL.tex

\end{document}

%% file: mysymbols.tex
\def\mmiss        {\mbox{$m^2_{\rm miss}$}\xspace}
\def\emiss        {\mbox{$E^*_{\rm miss}$}\xspace}
\def\eres        {\mbox{$E^*_{\pi}$}\xspace}
\def\pres        {\mbox{$p^*_{\pi}$}\xspace}

\def\plep       {\mbox{$p^*_{l}$}\xspace}
\def\plepmu       {\mbox{$p^*_{\mu}$}\xspace}
\def\plepe       {\mbox{$p^*_{e}$}\xspace}

\newcommand{\lumion}  {\mbox{211\invfb}}

\def\hhhnu {\ensuremath{\tau^- \rightarrow \pi^- \pi^+ \pi^- \nu_\tau}\xspace}
\def\hnu {\ensuremath{\tau^- \rightarrow \pi^- \nu_\tau}\xspace}

\def\eeltau {\ensuremath{e^+e^- \rightarrow l^+\tau^-}\xspace}
\def\eeetau {\ensuremath{e^+e^- \rightarrow e^+\tau^-}\xspace}
\def\eemtau {\ensuremath{e^+e^- \rightarrow \mu^+\tau^-}\xspace}

\def\amu {\ensuremath{e^+e^- \rightarrow \mu^+\tau^- (\hhhnu)}\xspace}
\def\ael  {\ensuremath{e^+e^- \rightarrow   e^+\tau^- (\hhhnu)}\xspace}
\def\pimu {\ensuremath{e^+e^- \rightarrow \mu^+\tau^- (\hnu)}\xspace}
\def\piel {\ensuremath{e^+e^- \rightarrow   e^+\tau^- (\hnu)}\xspace}

\def\amuxstr  {4.9}
\def\aelxstr  {13.2}
\def\pimuxstr {8.9}
\def\pielxstr {9.4}

\def\muxstr  {3.8}
\def\elxstr  {9.2}

\def\amubftr  {\ensuremath{4.3\times10^{-6}}\xspace}
\def\aelbftr  {\ensuremath{11.7\times10^{-6}}\xspace}
\def\pimubftr {\ensuremath{7.9\times10^{-6}}\xspace}
\def\pielbftr {\ensuremath{8.4\times10^{-6}}\xspace}

\def\mubftr  {\ensuremath{3.4\times10^{-6}}\xspace}
\def\elbftr  {\ensuremath{8.2\times10^{-6}}\xspace}

\def\ulimit {\ensuremath{90\%}\xspace}

\def\kk       {\mbox{\tt KK2F}\xspace}

\newcommand{\gevccgevcc}{\ensuremath{{\mathrm{\,Ge\kern -0.1em V^2\!/}c^4}}\xspace}

\newcommand{\evcc}{\ensuremath{{\mathrm{\,e\kern -0.1em V\!/}c^2}}\xspace}

%% file: authors_jul2006.tex
%
\author{B.~Aubert}
\author{M.~Bona}
\author{D.~Boutigny}
\author{F.~Couderc}
\author{Y.~Karyotakis}
\author{J.~P.~Lees}
\author{V.~Poireau}
\author{V.~Tisserand}
\author{A.~Zghiche}
\affiliation{Laboratoire de Physique des Particules, IN2P3/CNRS et Universit\'e de Savoie,
 F-74941 Annecy-Le-Vieux, France }
\author{E.~Grauges}
\affiliation{Universitat de Barcelona, Facultat de Fisica, Departament ECM, E-08028 Barcelona, Spain }
% ...
\collaboration{The \babar\ Collaboration}
\noaffiliation

%% file: acknow_PRL.tex
We are grateful for the excellent luminosity and machine conditions
provided by our \pep2\ colleagues, 
and for the substantial dedicated effort from
the computing organizations that support \babar.
The collaborating institutions wish to thank 
SLAC for its support and kind hospitality. 
This work is supported by
DOE
and NSF (USA),
NSERC (Canada),
IHEP (China),
CEA and
CNRS-IN2P3
(France),
BMBF and DFG
(Germany),
INFN (Italy),
FOM (The Netherlands),
NFR (Norway),
MIST (Russia), and
PPARC (United Kingdom). 
Individuals have received support from the 
A.~P.~Sloan Foundation, 
Research Corporation,
and Alexander von Humboldt Foundation.